\documentstyle[aps,preprint]{revtex}
\begin{document}
\title{Production of {\boldmath $A=6,7$} Nuclides in the
{\boldmath $\alpha+\alpha$} Reaction \\
and Cosmic Ray Nucleosynthesis}
\author{David~J.~Mercer$^{1,2}$\footnote{Present Address: NIS-5 Group,
LANL, Los Alamos NM 87545; Electronic address: mercer@lanl.gov},
Sam~M.~Austin$^{2,3}$\footnote{Electronic address:
austin@nscl.msu.edu}, J.~A.~Brown$^2$\footnote{Present address:
Physics Department, Millikin University, Decatur IL 62522},
S.~A.~Danczyk$^{2,3}$, S.~E.~Hirzebruch$^2$,
J.~H.~Kelley$^{2,3}$\footnote{Present address:  Department of
Physics, North Carolina State University,  Raleigh NC 27695},
T.~Suomij\"arvi$^{2,4}$ D.~A.~Roberts$^5$, T. P. Walker$^6$}

\address{$^1$Nuclear Physics Laboratory,
University of Colorado, Boulder, CO 80309-0446}
\address{$^2$National Superconducting Cyclotron Laboratory,
Michigan State University, East Lansing, MI  48824}
\address{$^3$Department of Physics and Astronomy,
Michigan State University,  East Lansing, MI  48824}
\address{$^4$Institute de Physique Nucl{\'e}aire, IN$_2$P$_3$-CNRS,
91406 Orsay, France}
\address{$^5$Department of Physics, University of Michigan, Ann Arbor, MI, 48109}
\address{$^6$Department of Physics, The Ohio State University,
Columbus OH 43210}
\date{\today}

\maketitle
\newpage

\begin{abstract}
Cross sections  for production of $^6$He, $^6$Li, $^7$Li, and
$^7$Be in the $\alpha+\alpha$ reaction were measured at bombarding
energies of 159.3, 279.6, and $619.8\;$MeV, and are found to
decrease rapidly with increasing energy. These cross sections are
essential for the calculation of the rate of nucleosynthesis of
the lithium isotopes in the cosmic rays and thereby play a key
role in our understanding of the synthesis of Li, Be, and B. The
results for $^6$Li differ significantly from the tabulated values
commonly used in cosmic-ray production calculations and lead to
lower production of $^6$Li.

\end{abstract}

\pacs{PACS number:}
\maketitle

\section{Introduction}
The origins of the light elements Li, Be, and B (LiBeB) differ
from that of the other nuclides.  Most elements are formed in
stars, but LiBeB are rapidly consumed by radiative capture
reactions in stellar centers and must therefore be synthesized in
cooler or more tenuous environments. It had been generally
accepted \cite{reeves70,MEN71,austin81,reeves94} that $^6$Li,
$^9$Be, $^{10}$B, and some $^{11}$B were made in the galactic
cosmic rays (GCR) by the interaction of fast GCR protons and
$\alpha$ particles with interstellar targets of carbon, nitrogen,
oxygen (CNO) or He (and vice-versa); $^2$H, $^{3,4}$He, and the
primeval abundance of $^7$Li were made in the Big Bang.

Recent measurements of abundances in metal-poor stars formed early
in the life of the galaxy challenge the details of the GCR
picture.  The abundance of $^9$Be rises linearly with the iron
abundance\cite{duncan92,steigman93}.  It has been argued (see
\cite{ramaty98,ramaty99} and references therein) that this means
that heavy cosmic rays (CNO) incident on interstellar hydrogen and
helium are responsible for the synthesis of LiBeB.
Others\cite{fields99} argue that the original process remains
viable. In any case the $\alpha+\alpha$ reaction plays a major
role in production of $^{6,7}$Li, particularly if light cosmic
rays are responsible for their synthesis\cite{steigman93}, since
the interstellar medium contains little CNO in the early galaxy.
Recent models \cite{ramaty99,fields99a} show that $^6$Li is
marginally produced in the observed quantity, so that accurate
estimates of the Li-producing reactions are required for a
rigorous test of these models.

Unfortunately, the $\alpha+\alpha$ cross sections\cite{read84} are
not known at high enough energies for such purposes---no data are
available for $^6$Li production above 200 MeV, although the cosmic
ray $\alpha$-particle flux remains strong beyond this energy
\cite{reeves94}. The predicted early-galaxy abundances of $^6$Li
can vary significantly (for example, by a factor of two in the
model of Ref. \cite{fields94}), depending on how the lower energy
cross sections are extrapolated to higher energy.

To provide the necessary data, we measured angular distributions
of $A=6,7$ ejectiles from the $\alpha + \alpha$ reaction  for alpha energies
between 159 and 620 MeV, and integrated these distributions to obtain the
total production cross sections for masses 6 and 7.  We find that the
cross sections fall rapidly, essentially exponentially, with
increasing bombarding energy, becoming small enough that $\alpha +
\alpha$ will not contribute significantly to cosmic ray nucleosynthesis
at energies above the measured range (to 620 MeV).

\section{Experimental Methods}

The cross sections at bombarding energies above 160 MeV
were expected to be small, leading to
potentially severe problems of background scattering and long
collection times if a traditional gas cell approach was used, as in
previous measurements\cite{KIN77,GLA78,ALA79,WOO85}.
To avoid these problems, we employed two novel techniques:
(1) the detectors were placed inside the target gas volume to
eliminate background scattering from target cell walls and windows, and
(2) a telescope system with a wide and continuous acceptance
in laboratory scattering angle was used to reduce data collection time.
Details of the target, detector telescopes, beam, and angle and acceptance
calibrations are given below.


The target was natural helium gas, $>$99.99\% pure, which filled
the NSCL 92-inch scattering chamber;  the gas pressure ranged from
370 to 408 torr (measured to $\pm 0.5\% $) and the temperature was
23$^\circ$C. A schematic of the chamber is shown in Fig.~1.  The
chamber housed a remotely operable turntable on which the
detectors were mounted, and the effective target was a cylinder of
gas extending approximately 135 cm upstream and 15 cm downstream
from the center of the turntable.
The entrance window was a 6.4 $\mu$m Havar foil which was
located 150 cm upstream from the turntable center and was
partially blocked from view of the detectors by lead shielding.
The chamber was evacuated on several occasions to check background
from the entrance foil. Only at the most forward laboratory
scattering angles ($\le 7^\circ$) were background subtractions
for window scattering necessary, and these subtractions
were smaller than 20\% of the total observed yield at any angle.

A target ladder was located at the center of the turntable, and
held secondary targets used for beam monitoring and calibration: a
47 mg/cm$^2$ carbon foil, a 0.5 mm diameter ``pinhole,'' a 1 mm
diameter vertical steel ``needle," a 2 mm thick aluminum target
with a 20 mm diameter hole, and a scintillator.  This ladder was
lowered completely out of the beam path during data collection
with the primary (helium) target.


Beams of $\alpha$ particles were provided by the NSCL K1200
Cyclotron at nominal energies of 160, 280, and 620 MeV.
The lowest energy was chosen to allow direct comparison with
results of \cite{GLA78}, and the higher energies were chosen to
span the region of astrophysical interest. All beams were
fully stripped upon arrival at
the scattering chamber. Up to $1.2$ MeV is deposited in the entrance
foil and target gas prior to the reaction, depending on the
initial energy and point of scatter. The average
bombarding energies, accounting for these losses
and the 0.2 MeV uncertainty in the measured energy,
were $159.3\pm0.5$ MeV, $279.6\pm0.4$ MeV, and $619.8\pm0.3$ MeV.
The beam spot was checked by periodic insertion of two
scintillators viewed by TV cameras, located at the turntable
center and 170 cm upstream of turntable center.  The beam radius at the
turntable was  $\leq 2\;$mm and the calculated half-angle divergence
$\leq 0.5^\circ$. Possible beam halo was monitored by
periodically inserting the aluminum ``hole" target and looking
for scattered particles; no significant halo was observed.
The unscattered beam was collected in a vacuum-isolated Faraday
cup 2 meters downstream from the scattering chamber, and
integrated beam current was found with $\pm$5\% uncertainty using a
BIC Model 1000 current integrator.


Two detector telescopes were fixed to the turntable at beam
height, one on each side of the beam.  Each telescope consisted of
a stack of two $1\;$cm tall $\times$ $9 \;$cm wide charge-division
position-sensitive silicon detectors (PSDs) and one 3 cm thick
CsI(Tl) scintillator viewed by photodiodes, as shown in Fig.~2. In
telescope 1 the PSDs had thicknesses of 320 $\mu$m (front) and
1000 $\mu$m (back), and were located 300 and 367 mm from the
turntable center.  In telescope 2, the PSDs had thicknesses of 300
$\mu$m (front) and 480 $\mu$m (back), and were located 300 and 363
mm from the turntable center, 80$^\circ$ clockwise from telescope
1. Standard  electronics were used for signal amplification and
data acquisition.

With the turntable centered, each telescope viewed a laboratory
scattering angle range of 7$^\circ$--60$^\circ$ simultaneously.
Five different turntable settings were used during data
collection;  at a given scattering angle this provided independent
measurements with a variety of solid-angle acceptance conditions.
The turntable settings were centered (as shown in Fig.~1), rotated
$\pm 10^\circ$ from the centered position, and rotated $\pm
20^\circ$.  The smallest laboratory angle observed was $4^\circ$,
seen only by the telescope rotated closest to the beam for
$\pm20^\circ$ settings.

The telescopes provided distinct particle identification for
$A\leq 7$ ejectiles from characteristic $\Delta E$-$E$ signatures
in the PSDs. In some kinematic situations the ejectile penetrated
both PSDs and stopped in the CsI(Tl) scintillator; in this case
the energy deposited in the scintillator was also used to aid
particle identification. The mass resolution was about 0.4 amu
FWHM, sufficient to distinguish $^6$Li and $^7$Li.  The energy
resolution was sufficient to distinguish elastically scattered
$^4$He from $^4$He produced in inelastic channels such as
$^4$He$(\alpha,tp)^4$He ($Q=-19.8\;$MeV); however, discrimination
between the $^4$He$(\alpha,d)^6$Li channel ($Q=-22.4\;$MeV) and
the $^4$He$(\alpha,pn)^6$Li channel ($Q=-24.6\;$MeV) was
unreliable, particularly at the higher energies. It was not
possible to distinguish the particle-stable first excited state in
$^7$Li or $^7$Be (0.48 MeV and 0.43 MeV, respectively) from the
ground state. The first excited state in $^6$Li (2.18 MeV) decays
immediately to $\alpha+d$ and does not contribute significantly to
$^6$Li production. The second excited state in $^6$Li (3.56 MeV)
is particle-stable, but it was not possible to distinguish it from
the ground state. Since the primary concern of the present
experiment is total production cross sections, these resolution
limitations are not important.


For a valid event, both PSDs in a telescope must detect the
scattered particle.  The impact position on each PSD is inferred
using standard charge-division techniques; from these impact
positions and the known spacing between the PSDs it is possible to
derive the scattering angle $\theta_{\rm lab}$ and the scattering
position $z$ along the beam axis. The scattering angle is needed
for computing the differential cross sections
$d\sigma/d\Omega_{\rm lab}$, and the $z$ position is needed to
exclude particles that may have scattered in the Havar entrance
foil.  Angle calibration is performed by stepping each telescope
through a pinhole-collimated $\alpha$ beam.  The calibration was
tested by reconstructing the tracks of particles scattered from a
``needle" target protruding into the beam and from four $50\mu$m
Kapton foils that could be inserted at various locations along the
beam path. Angular resolution was about 1.2$^\circ$ FWHM and was
not degraded significantly when the chamber was filled with
helium.

Differential cross sections are found from the observed yields
according to
\begin{equation}
\frac{d\sigma(\theta)}{d\Omega_{\rm lab}} =\frac{Y(\theta)}{N_{\rm
beam}\; f_{\rm live}\; \rho\; A(\theta)}    ,
\end{equation}
where $Y(\theta)$ is the observed number of particles at the
laboratory scattering angle $\theta $. $N_{\rm beam}$ is the
number of incident particles, $f_{\rm live}$ is the detection
system live fraction, and $\rho$ is the target density
(nuclei/cm$^3$).  The detector acceptance is denoted by
$A(\theta)$ and has units of cm. $A(\theta)= \Delta\Omega x
\epsilon T$, where $\Delta\Omega(\theta)$ is the effective solid
angle, $x(\theta)$ is the effective length of the target,
$\epsilon$ is the detector efficiency, and $T$ is the transmission
fraction of scattered particles.

$A(\theta)$ depends primarily on the geometry  of the detector and
the incident beam. It is determined by comparing our observed
yields of elastically scattered $^4$He at 160 MeV with the
laboratory cross sections reported by Nadasen {\it et al.}
\cite{NAD78} at a similar energy. The data of \cite{NAD78} have
better angular resolution than the present experiment, and had to be
slightly degraded before the comparison. By solving equation (1)
for $A(\theta)$ in one degree (laboratory frame) bins, we calibrate the
acceptance for all angles of interest. The process is repeated for
each of the five turntable settings.

For practical purposes the $A(\theta)$ calibration is independent
of energy and particle species. Energy or species dependence can
occur because of transmission losses through the target gas or
reactions in the Si or CsI detectors.  The former is always less
than 1\% in the kinematic region of interest. Reaction losses in
the detectors can cause a loss of a few percent of detector
efficiency in the worst cases.  However, losses in the $A=6,7$
efficiency tend to  offset the losses in the efficiency for $^4$He
particles used for calibration, and the net effects are much
smaller than the quoted uncertainty. Although different particles
have different cutoff angles (this occurs when the the particle
energy is insufficient to penetrate the first PSD), the cutoff
angles are all in the backward c.m. hemisphere, which is not used
in the analysis.

Several quality checks were applied to the $A(\theta)$
calibration. (1) The differential cross sections derived from the
five different turntable angles were compared, and were found to
be consistent. (2) $A(\theta)$ was compared with (somewhat
simplified) Monte Carlo simulations of the detector geometry and
were found to agree within 12\%. (3) Our $A=7$ differential cross
sections at 160 MeV are in excellent agreement with those by
Glagola, {\it et al.} \cite{GLA78}, which lends confidence to our
methods. We estimate a systematic uncertainty of 8\%, which is
dominated by the absolute uncertainty in the results of Ref.
\cite{NAD78}.

\section{analysis and results}

The differential and total cross sections were measured for the
four reactions $^4$He$(\alpha,pp)^6$He,
$^4$He$(\alpha,\{d,pn\})^6$Li, $^4$He$(\alpha,p)^7$Li, and
$^4$He$(\alpha,n)^7$Be. Because the target and projectile are
identical, reaction products are distributed symmetrically about
$90^\circ$ c.m.; it is then sufficient to determine yields only
for $\theta_{\rm c.m.}\leq 90^\circ$. Detector acceptance for
events from the backward c.m. hemisphere was rather low because
the low energy ejectiles stopped in the 300 $\mu$m or 320 $\mu$m
PSDs.

In all four reactions of interest, the kinematics are ``folded''
so that each laboratory angle corresponds to two c.m. scattering
angles.  For the two-body final states leading to $^7$Li and
$^7$Be, it is a simple matter to distinguish between the two c.m.\
angles, because a larger ejectile energy is always associated with
the more forward angle. There is an added complication for $^6$Li
and $^6$He due to the three-body exit channels $^6$He$+p+p$ and
$^6$Li$+p+n$.  We determined a locus of ejectile energies
corresponding to 90$^\circ$ c.m., as a function of laboratory
scattering angle, from the kinematics result
\begin{equation}
p_{\rm 6Li}\cos(\theta) = \gamma m_{\rm 6Li}V_{\rm c.m.}    ,
\end{equation}
where $p_{\rm 6Li}$ is the momentum of $^6$Li and $V_{\rm c.m.}$ is the
 velocity of the center of mass;
all quantities are measured in the lab. The result is shown in
Fig.~3. All $^6$Li ejectiles observed with energies above this
locus were attributed to the forward c.m.\ hemisphere, and those
with less energy were attributed to the backward hemisphere.  A
similar procedure was used for the $^6$He data.  One can, in
principle,  distinguish the $^4$He$(\alpha,d)^6$Li reaction from
the $^4$He$(\alpha,\{pn\})^6$Li reaction, as it is restricted to a
band labeled ``0 MeV" in Fig.~3. This was done in Refs.\
\cite{GLA78,WOO85}, but our energy resolution made it impossible;
this is not a limitation for use of these cross sections in
calculations of cosmic ray nucleosynthesis .

\subsection*{A. Differential cross sections}

The differential cross sections (forward c.m.\ hemisphere only)
for $^6$Li and $^6$He are shown in Fig.~4. The cross sections for
$^6$Li are forward peaked at all three energies. The cross
sections for $^6$He are also forward peaked and are roughly $1/8$
as large as those for $^6$Li. It was not possible to measure the
cross section for $^6$He at 620 MeV due to background
contamination from $^{3,4}$He ejectiles.  We established that the
$^6$He cross section at 620 MeV is less than 25\% of the cross
section for $^6$Li at this energy.


The differential cross sections (forward c.m.\ hemisphere only)
for $^7$Be and $^7$Li are shown in Fig.~5. At 160 MeV the cross
sections for both isotopes are forward-peaked with a minimum at
90$^\circ$ c.m.;  they are in excellent agreement with the measurements
of Ref. \cite{GLA78} at a similar energy.
The cross sections at 280 MeV are two orders
of magnitude smaller, approaching the limit of our experimental
method, and the minima at 90$^\circ$ disappear. At 620 MeV
the cross sections were very small: fewer than two dozen
possible $^7$Li and $^7$Be events were observed, roughly consistent
with background. An upper limit is obtained at 620 MeV.

\subsection*{B. Total cross sections}


Total cross sections for $A=6$ production were found by extrapolating
the differential cross sections to zero
degrees, integrating over all laboratory angles corresponding to
$\theta_{c.m.}\leq 90^\circ$,
and doubling to account for the $90^\circ-180^\circ\;$c.m.\ yield.
The extrapolation to zero degrees is based on a linear fit through
the three smallest-angle points ($\theta_{\rm lab}=4.5, 6.5,$ and
$8.5^\circ$). The extrapolated cross section $0-4.5^\circ$ typically
accounts for $10-20$\% of the total cross section, and
up to one third of the random uncertainty.

Total cross sections for $A=6$ production are shown in Fig.~6 and
Table I. For comparison, the cross sections from previous
measurements in the 60-200 MeV range are also included.  Errors
given for the present experiment include the statistical,
extrapolation, and normalization (8\%) uncertainties, all added in
quadrature.

The measured cross sections for $^6$He and $^6$Li differ from the
values of Glagola {\it et al.}~\cite{GLA78} near 160 MeV by about
a factor of two. However, the present results agree with the
reanalysis of the Glagola, {\it et al.} data by Mercer, Austin,
and Glagola \cite{MER96}. For both nuclides the total cross
sections decrease rapidly with increasing energy.  The solid lines
in Fig.~6 are weighted exponential fits through all points shown,
and have the functional forms (with the bombarding energy,
$E_\alpha $, in MeV and the slope for $^6$He taken to be identical
to that obtained for $^6$Li)
\begin{equation}
\sigma_{\rm 6Li}=66\exp{(-0.0159 E_\alpha)}\; \rm mb    ,
\end{equation}
\begin{equation}
\sigma_{\rm 6He}= 9.3\exp{(-0.0159 E_\alpha)}\; \rm mb    ,
\end{equation}
The exponential falloff is shallower than that suggested by Woo
{\it et al.}\cite{WOO85}, who report fits proportional to
$\exp(-0.025E_\alpha)$ for both $^6$He and $^6$Li.  The difference
is understandable because the reanalyzed lower energy data
\cite{MER96} were not available to Woo {\it et al.}.  The
interpolated line for $^6$He lies somewhat above the upper limit
from \cite{WOO85} at 198 MeV, but in general the data form a
consistent set.  The cross section for $^6$Li at 620 MeV lies
significantly above the fitted exponential.


Total cross sections for $^7$Li and $^7$Be are found by
extrapolating the differential cross sections to zero degrees,
integrating from $0-90^\circ$ in the c.m. frame, and doubling, as
in the $A=6$ analysis. Results are shown in Fig.~7 and in Table II
along with other measurements from 60 to 600 MeV. Our result at
159.3 MeV is in excellent agreement with those of Glagola
\cite{GLA78} near 160 MeV; an exponential describes the entire
data set very well. Our experiment yields a significantly tighter
upper bound for the $^7$Be cross section near 600 MeV than that
given in \cite{YIO77}, and a new limit for $^7$Li.  As expected,
since the channels are isospin mirrors, the $^7$Li and $^7$Be
cross sections are similar in size and energy dependence. Weighted
exponential fits yield:
\begin{equation}
\sigma_{\rm7Li}=299\exp(-0.0362 E_{\alpha})\;\rm mb    ,
\end{equation}
\begin{equation}
\sigma_{\rm7Be}=208\exp(-0.0343 E_{\alpha})\;\rm mb    .
\end{equation}
These results are quite similar to that reported in
Ref.~\cite{WOO85}: $260\exp(-0.035 E_{\alpha})\rm mb$.

Since $^6$He decays to $^6$Li by $\beta^-$ emission, with a
half-life of about 807 msec, the $^4$He$(\alpha,pp)^6$He and
$^4$He$(\alpha,\{d,pn\})^6$Li reactions are both a source of
$^6$Li produced in the cosmic rays.
 Therefore, the sum of
$^6$Li and $^6$He cross sections at each energy is also given in
Table I.  In cases where $^6$He measurements are not available,
the sum includes an estimate for the $^6$He contribution, equal to
12\% of the $^6$Li cross section (103.0 and 619.7 MeV) or equal to
the reported upper limit (0.2 mb) for $^6$He at 198.4 MeV.

In Fig.~8 we show the mass--6 cross sections, the sum of the cross
section for $^6$Li and $^6$He. As was already clear from the fits
shown in Fig.~6, an exponential fit would lie significantly (about
three standard deviations) below the point at 620 MeV.  The fit
shown includes an energy independent cross section and describes
the data well; it should be useful for applications.  It would be
desirable to use a form that more accurately reflects possible
physical processes at high energy, but the data is sufficient to
fix only one parameter beyond the low energy exponential; adding a
constant cross section is the simplest choice. The resulting
constant cross section is much smaller than the uncertainty in the
low energy cross sections.

The sum of the $^4$He$(\alpha,p)^7$Li and $^4$He$(\alpha,p)^7$Be
cross sections is also given in Table II for each energy.  Since
$^7$Be decays by electron capture to $^7$Li with a half-life of
53.3 days, these are the relevant cross sections for calculation
of $^7$Li production in cosmic rays. Finally, Fig.~8 shows the sum
of the $^7$Li and $^7$Be cross sections, an exponential fit to the
data, and a fit including a constant cross section.

The exponential or exponential-plus-constant cross section forms
shown in Fig. 8 provide a convenient description of the data for
use in applications. For the total mass-6 and mass-7 yields these
are
\begin{equation}
\sigma_{mass-6}=0.014+75\exp{(-0.0159 E_\alpha)}\; \rm mb    ,
\end{equation}
\begin{equation}
\sigma_{mass-7}=0.005 + 514\exp(-0.0354 E_{\alpha})\;\rm mb    ,
\end{equation}
\begin{equation}
\sigma_{mass-7}=510\exp(-0.0354 E_{\alpha})\;\rm mb    .
\end{equation}

\section{Summary and discussion}

With these new cross sections for production of $A=6,7$ by the
$\alpha+\alpha$ reaction, it is possible to calculate the amounts
of $^6$Li and $^7$Li produced in early-galaxy cosmic rays more
accurately.  The need to extrapolate cross sections to high
energies is no longer a significant factor in the uncertainty.
Because of the smaller cross sections obtained here for mass-6 the
the production of $^6$Li will be significantly smaller than would
be obtained with cross sections from the the summary of Read and
Viola\cite{read84}.

To illustrate these points, we use a cosmic ray spectrum peaked
around 200 MeV/nucleon (specifically, Fig. 2b of
Ref.\cite{fields94}, the curve labeled $\Lambda = 10$ g/cm$^2$.
The product of this spectrum and various cross sections was
integrated over energy. A comparison of results using our cross
sections and those of  Read and Viola\cite{read84} is useful,
because most calculations of cosmic ray nucleosynthesis have used
the Read-Viola cross sections. Both an exponential fit (not shown)
and the exponential plus background fit (Eq. 7) shown in Fig.~8
yield about 50\% of the $^6$Li obtained using the cross sections
of Read and Viola. For $^7$Li the differences are relatively
small, because the reaction cross sections are already quite small
by 200 MeV, and because upper limits for $^7$Be at higher energies
had been reported. For the cosmic ray spectrum employed here (and
using the Read-Viola cross sections for $E<70$ MeV) the production
rates of $^7$Li and $^6$Li are nearly equal, with $^7$Li
production larger by about 10\%.

Detailed calculations of the effects of the new cross sections and
the suggested renormalizations of lower energy cross
sections\cite{MER96} will be reported at a later date
\cite{WAL96}. In general they will result in a significant
reduction of the calculated cosmic ray production of $^6$Li by
$\alpha+\alpha$ reactions. The reduction is dependent on the
cosmic ray spectrum and will be largest when it has a large
component above 200 MeV. The energy dependence of the cross
sections at high energies, that is, the expected limit of the
observed exponential behavior, remains an issue. We are not aware
of detailed studies of this phenomenon. In the case of mass-6, a
deviation from exponential behavior is required by the data, and
we have assumed a constant value at higher energies. In the case
of mass-7, there is no convincing evidence for such a deviation.
However, given the deviation seen for mass-6, we have provided for
mass-7, as  alternatives, pure exponential and
constant-plus-exponential fits to the data. For the cosmic ray
spectrum we have chosen, the difference between the two
assumptions affects the mass-7 yield at only the 0.4\% level.

The production of $^6$Li and $^7$Li  by $ \alpha+\alpha $
reactions will play a vital role in reaching an understanding of
the synthesis of LiBeB and the nature of the cosmic rays.  The
critical questions are whether the $^7$Li observed in old
metal-poor stars is that produced in the Big Bang, or whether Big
Bang $^7$Li has been affected by production in cosmic rays and
destruction by stellar processing. Recent data indicate that the
amount of Li increases with time, presumably an indication of
cosmic ray production.  However, our data bear more strongly on
the possibility of stellar destruction of $^{6,7}$Li.  Since
$^6$Li is more fragile than $^7$Li, its survival in a star can be
used to limit the amount of $^7$Li depletion.

In order to use $^6$Li in this way, the amount of $^6$Li formed
must be estimated from a model for cosmic ray nucleosynthesis
(very little $^6$Li is made in the Big Bang) and then compared to
its observed abundance.  The resulting $^6$Li destruction might
then be used to estimate the amount of $^7$Li destruction and
eventually, the primordial abundance of $^7$Li. In recent models
(see for example\cite{ramaty99,fields99a}) the production of
$^6$Li is marginally sufficient or too small, even assuming none
has been destroyed during stellar evolution. The downward changes
in the predictions that will result from the present measurements
may, therefore, have important consequences.

\acknowledgements
 We would like to thank P. Danielewicz, B. D.
Fields, R. Ramaty, and V. E. Viola for useful discussions; M.
Hellstrom, R. A. Kryger, J. S. Winfield, and the staff of the
National Superconducting Cyclotron Laboratory for assistance with
various aspects of the data collection; and R. J. Peterson, R. A.
Ristinen, and C. J. Gelderloos for useful discussions concerning
analysis. This work was supported by grants from the National
Science Foundation and the U.S. Department of Energy.

\begin{table}[htb]
\caption{Total cross sections (mb) for the $^4$He$(\alpha,pp)^6$He
and $^4$He$(\alpha,\{d,pn\})^6$Li (g.s.+3.56) reactions as a
function of bombarding energy.  Their sum is also given for
convenience, with estimates made for the $^6$He contribution as
described in the text where data are not available.  The errors
for the present experiment include an 8\% uncertainty in the
normalization, added in quadrature.}

\begin{tabular}{cccc}
Energy& & &\\(MeV)&$^6$He & $^6$Li & $^6$He$+^6$Li\\ \hline

61.5    &$\rm 1.7\pm0.2^a$&$\rm 21.5\pm2.2^b$&$23.2\pm2.2$\\
 80.8 &$\rm 2.0\pm0.6^b$&$\rm 18.9\pm1.4^b$&$20.9\pm1.5$\\ 103.0
&$-$&$\rm 12\pm1^c$&$14\pm2$\\ 118.9   &$\rm 1.2\pm0.3^b$&$\rm
10.5\pm0.6^b$&$11.7\pm0.7$\\ 139.2   &$\rm 1.1\pm0.3^b$&$\rm
8.4\pm0.5^b$&$9.5\pm0.6$\\ 158.2   &$\rm 0.8\pm0.2^b$&$\rm
5.2\pm0.3^b$&$6.0\pm0.4$\\ {\bf 159.3} &$\bf 0.79\pm0.08$$^{\rm
d}$ &$\bf 5.3\pm0.4$$^{\rm d}$ &$\bf 6.1\pm0.5$\\ 198.4 &$\rm
<0.2^e$&$\rm 3.4\pm0.8^e$&$3.6\pm0.8$\\ {\bf 279.6} &$\bf
0.11\pm0.03$$^{\rm d}$ &$\bf 0.64\pm0.08$$^{\rm d}$ &$\bf
0.75\pm0.09$\\ {\bf 619.7} &$-$ &$\bf 0.015\pm0.004$$^{\rm d}$
&$\bf 0.018\pm0.004$\\
\end{tabular}
$^{\rm a}$Reference \cite{GLA78}.\\
$^{\rm b}$Reference \cite{GLA78}
renormalized according to \cite{MER96}.\\
$^{\rm c}$Reference \cite{ALA79}.\\
$^{\rm d}$Present experiment (boldface).\\
$^{\rm e}$Reference \cite{WOO85}.
\end{table}

\begin{table}[htb]
\caption{Total cross sections (mb) for the $^4$He$(\alpha,p)^7$Li
(g.s.+0.48) and $^4$He$(\alpha,n)^7$Be (g.s.+0.43) reactions as a
function of bombarding energy.  Their sum is also given. The
errors in the present results include an 8\% uncertainty in the
normalization, added in quadrature. Upper limits from the present
work are at the one standard deviation level.}

\begin{tabular}{cccc}
Energy & & &\\ (MeV)& $^7$Li & $^7$Be & $^7$Li$+^7$Be\\ \hline
61.5    &$\rm 33.2\pm2.7^a$&$\rm 23.1\pm2.6^a$&$56.3\pm3.7$\\ 80.8
&$\rm 16.8\pm1.1^a$&$\rm 15.6\pm1.7^b$&$32.4\pm2.0$\\ 103.0 &$\rm
6\pm1^c$&$\rm 6\pm1^c$&$12\pm1.4$\\ 118.9   &$\rm
4.0\pm0.3^a$&$\rm 3.6\pm0.3^b$&$7.6\pm0.4$\\ 139.2   &$\rm
2.0\pm0.2^a$&$\rm 1.8\pm0.2^b$&$3.8\pm0.3$\\ 158.2   &$\rm
0.95\pm0.08^a$&$\rm 0.83\pm0.07^b$&$1.78\pm0.11$\\ {\bf 159.3}
&$\bf 1.00\pm0.10$$^{\rm d}$ &$\bf 0.87\pm0.09$$^{\rm d}$ &$\bf
1.87\pm0.17$\\ 198.4   &$\rm 0.25\pm0.06^e$&$\rm
0.35\pm0.08^e$&$0.60\pm0.10$\\ {\bf 279.6} &$\bf
0.028\pm0.014$$^{\rm d}$ &$\bf 0.022\pm0.009$$^{\rm d}$ &$\bf
0.050\pm0.017$\\ 400.0   &$-$&$\rm <0.02^f$&$-$\\ 600.0 &$-$&$\rm
<0.014^f$&$-$\\ {\bf 619.7} &$\bf <0.004$$^{\rm d}$ &$\bf
<0.003$$^{\rm d}$ &${\bf <0.006}$
\end{tabular}
$^{\rm a}$Reference \cite{GLA78}.\\
$^{\rm b}$Ref.\ \cite{GLA78}
renormalized according to \cite{MER96}.\\
$^{\rm c}$Reference \cite{ALA79}.\\
$^{\rm d}$Present experiment (boldface).\\
$^{\rm e}$Reference \cite{WOO85}.\\
$^{\rm f}$Reference \cite{YIO77}.
\end{table}

\begin{figure}[dum]
\caption{Schematic of the experimental setup. The scattering
chamber had a volume of 13 m$^3$, and was filled with
$^{\rm nat}$He gas during normal data collection.  Two detector
telescopes were mounted on a turntable inside the chamber,
viewing an effective target region approximately 150 cm long.
Several reconstructed $^7$Li ejectile tracks are shown.}
\end{figure}

\begin{figure}[dum]
\caption{Detail of a telescope, consisting of two Micron, Inc.\
Model TT position-sensitive detectors and a CsI(Tl) scintillator
crystal; for more details see the text.  An example ejectile track
is shown.}
\end{figure}

\begin{figure}[dum]
\caption{Kinematics for the $^4$He$(\alpha,\{d,pn\})^6$Li reaction
at a bombarding energy of 280 MeV.  The dot-dash curve indicating
90$^\circ$ in the center of momentum frame was determined as
described in the text. The two-body $^6$Li$+d$ exit channel
follows the curve labeled ``0". The other curves are for the
three-body $^6{\rm Li}+p+n$ exit channel, approximated by assuming
a deuteron with pseudo-excitation energy up to 100 MeV. Solid
portions of the curves indicate the region of kinematic acceptance
of our detectors. These curves were not used in the analysis and
are given for orientation.} \end{figure}

\begin{figure}[dum]
\caption[dum]{Laboratory differential cross sections for the
$^4$He$(\alpha,\{ d,pn \} )^6$Li and $^4$He$(\alpha,pp)^6$He
reactions from the present experiment. Error bars show statistical
uncertainty. }
\end{figure}

\begin{figure}[dum]
\caption[dum] {Differential cross sections for the
$^4$He$(\alpha,n)^7$Be and $^4$He($\alpha,n)^7$Li reactions.
Squares and circles are from the present experiment, and
$\times$'s are from Ref. \cite{GLA78}.}
\end{figure}

\begin{figure}[dum]
\caption[dum]{Total cross sections for $^4$He$(\alpha,\{ d,pn \}
)^6$Li (squares) and $^4$He$(\alpha,pp)^6$He (circles). Solid
symbols are from the present experiment, and the open symbols are
from previous work as summarized in Table I. The lines are
exponential fits as described in the text (Eqs. 3, 4).}
\end {figure}

\begin{figure}[dum]
\caption[dum]{Total cross sections for $^4$He($\alpha,n)^7$Be
(circles) and $^4$He($\alpha,n)^7$Li (squares). All $^7$Be values
are multiplied by 10 for clarity. Solid symbols are from the
present experiment, and the open symbols are from previous work as
summarized in Table II. The lines are exponential fits as
described in the text (Eqs. 5, 6).}
\end{figure}

\begin{figure}[dum]
\caption[dum]{Total cross sections for mass-6 and mass-7.  The
fits are described in the text (Eqs. 7-9).}
\end{figure}

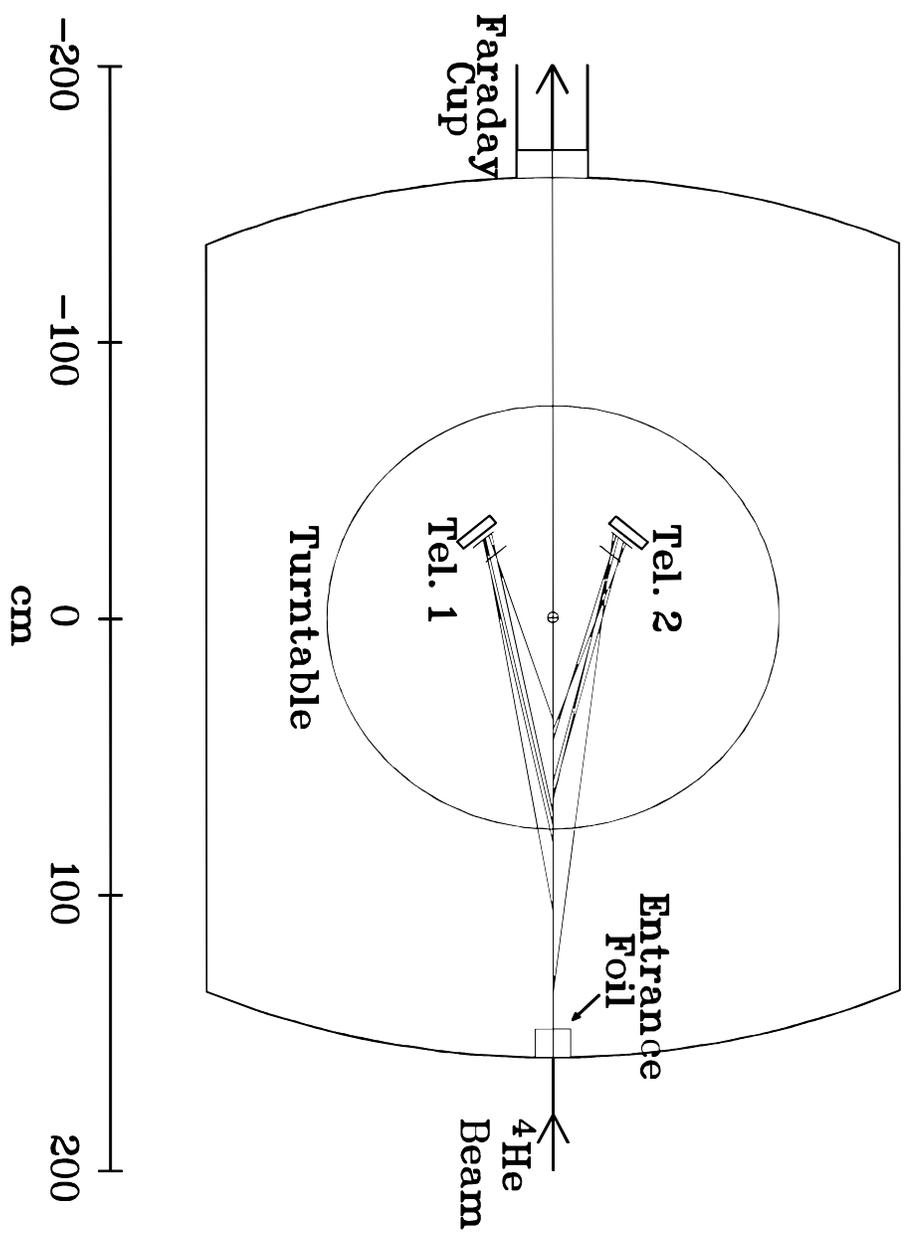
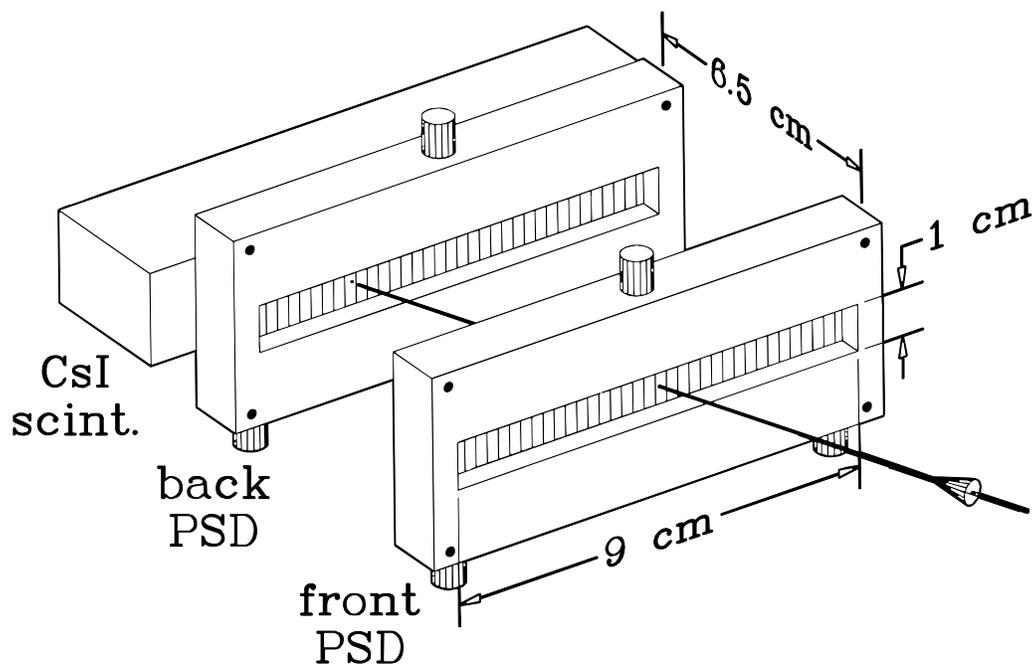
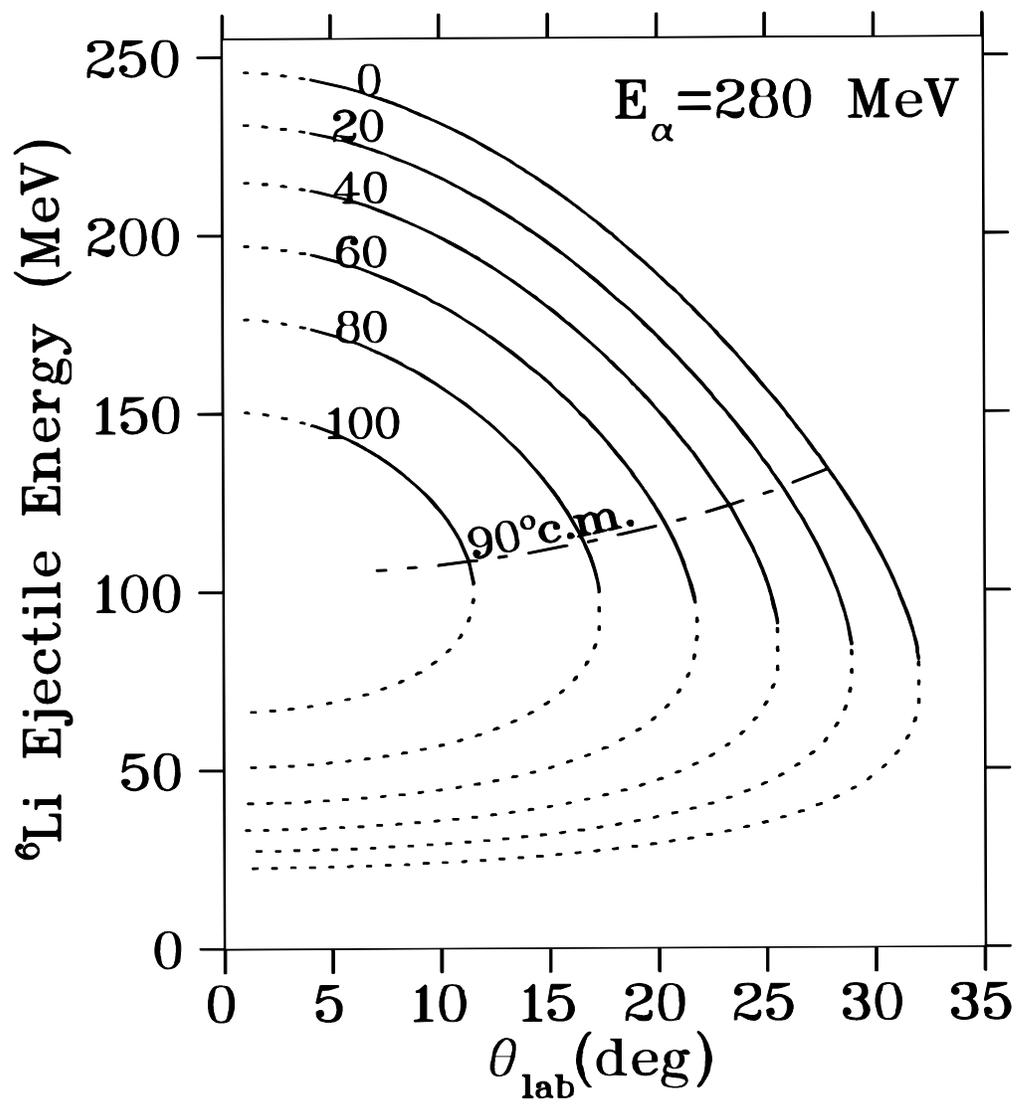
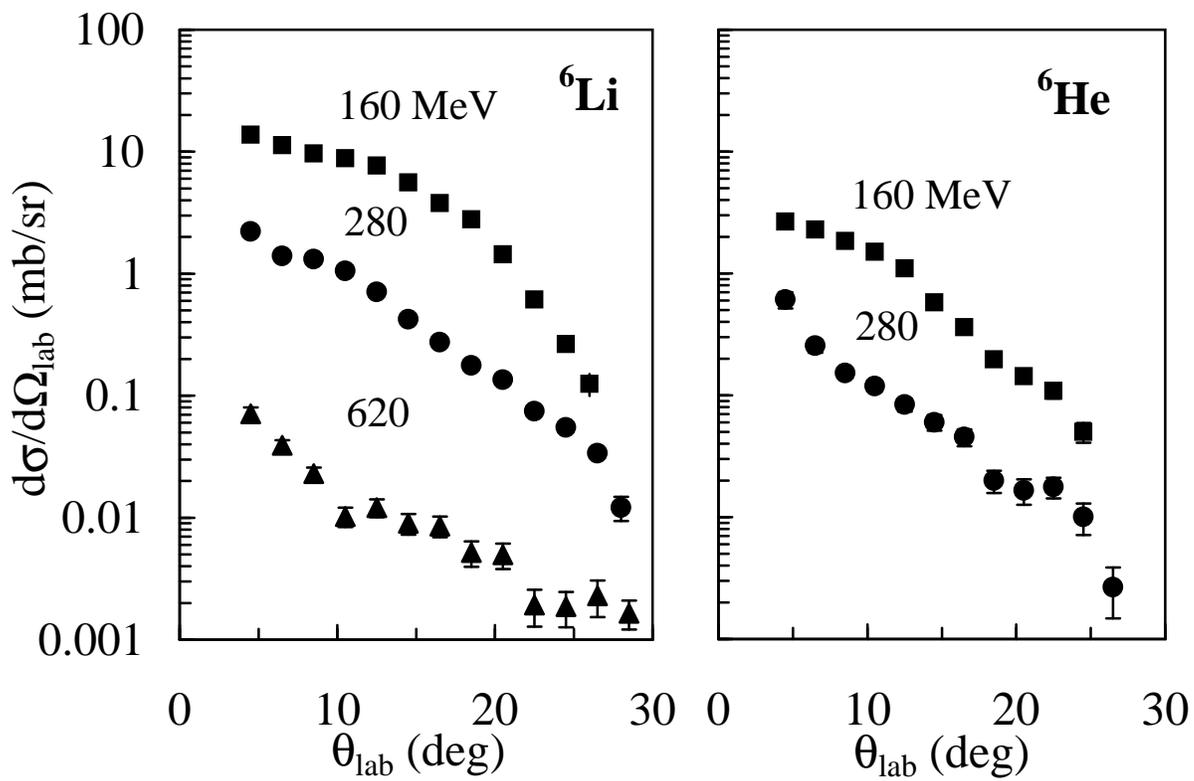
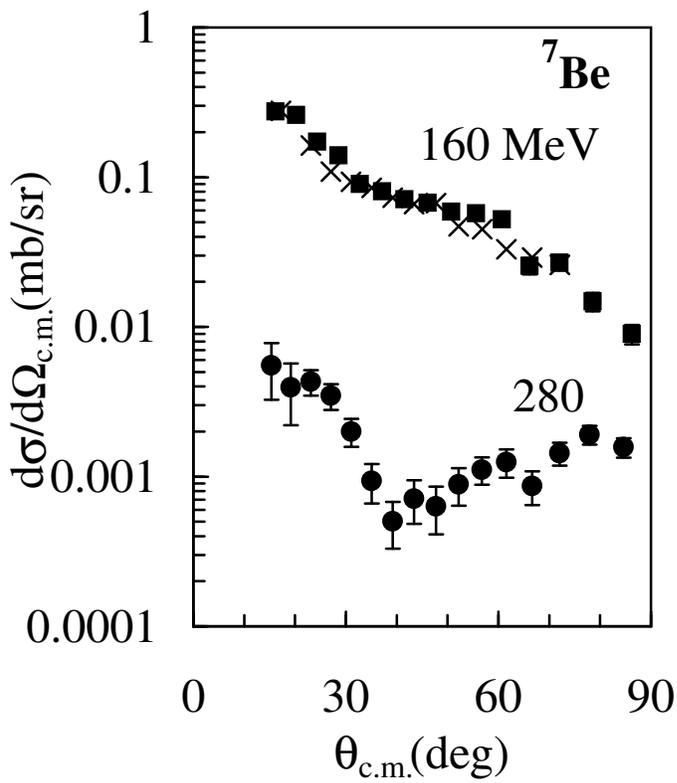
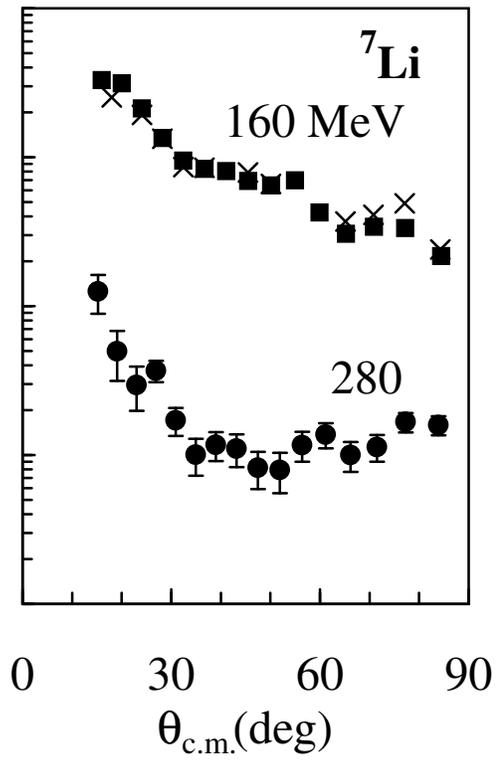
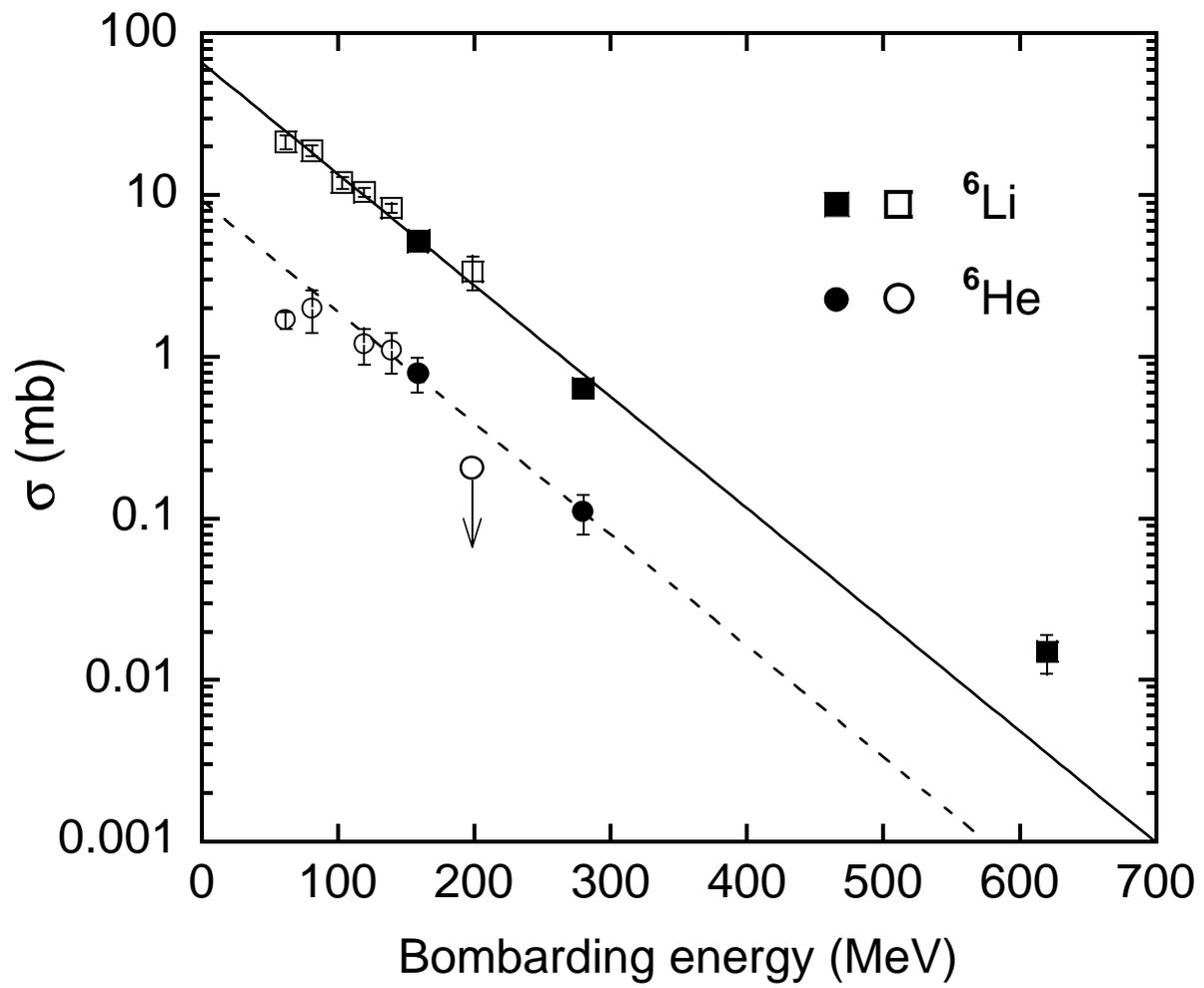
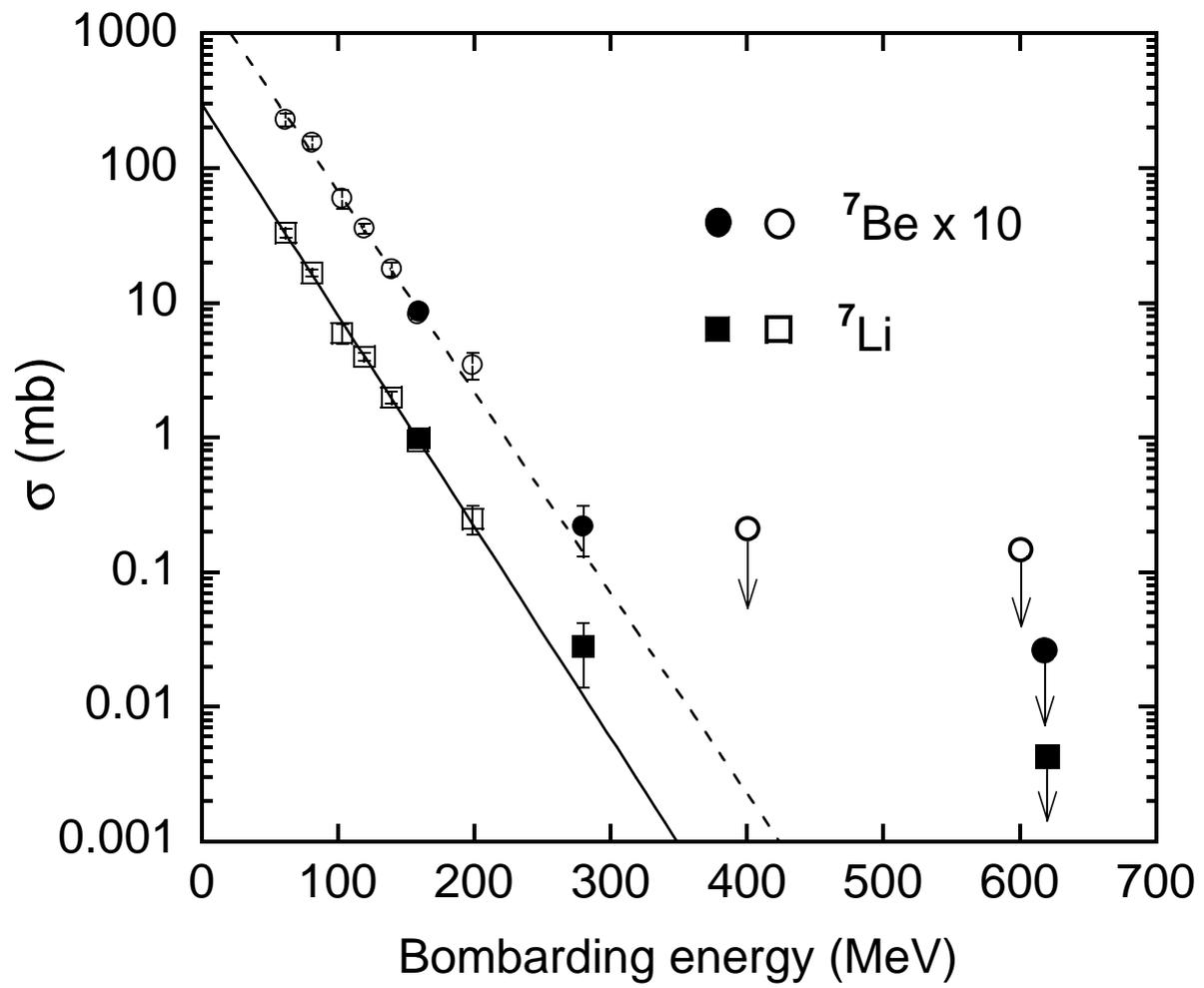
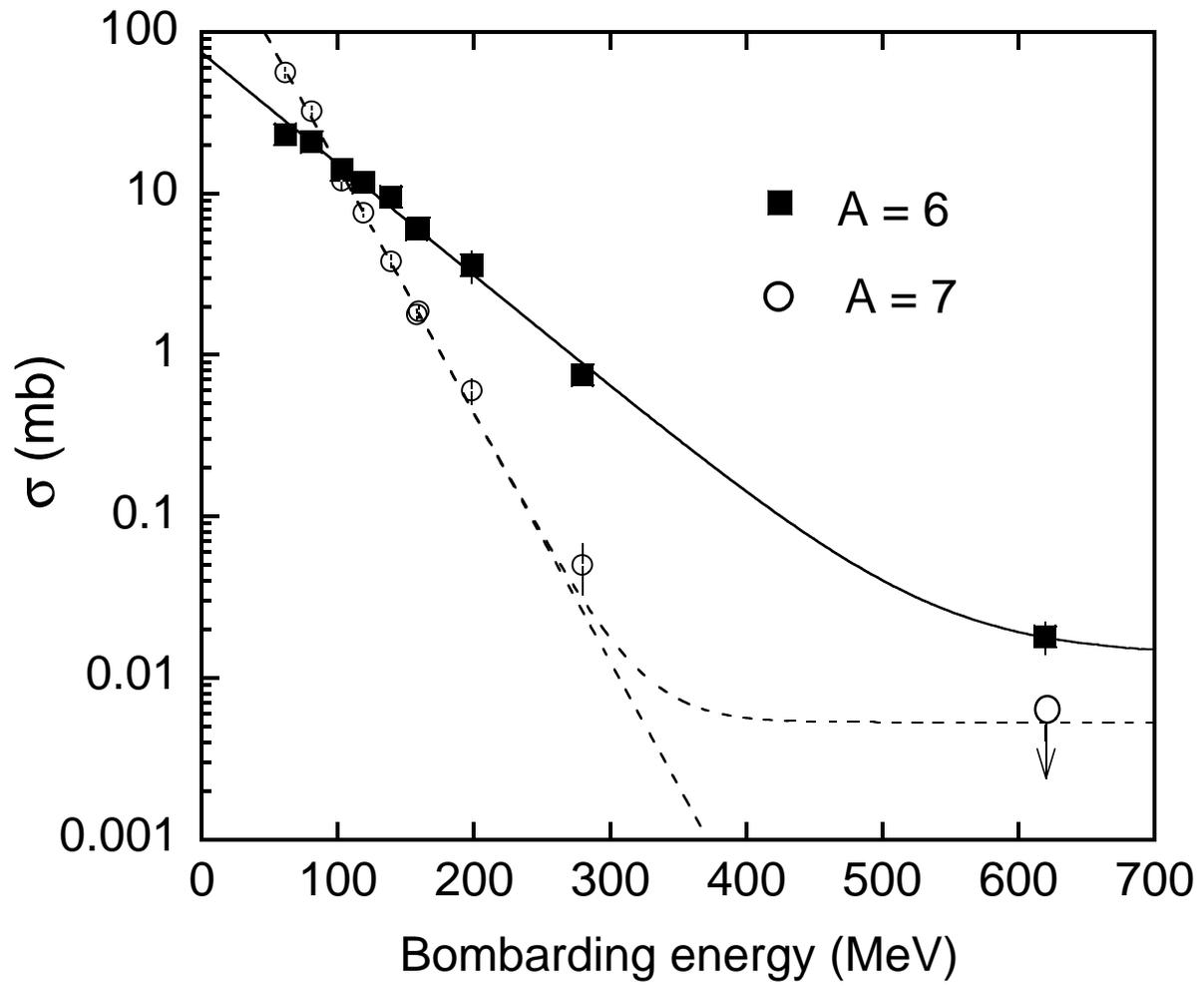
\end{document}